\documentclass[article]{jss}

\usepackage{orcidlink,lmodern}
\usepackage{amsmath}
\usepackage{cleveref}

\usepackage{bm}
\renewcommand{\vec}{\bm}
\newcommand{\like}{L}
\newcommand{\ev}{Z}
\newcommand{\params}{\bm{\theta}}
\newcommand{\dint}{\,\text{d}}
\newcommand{\nlive}{n_\text{live}}
\newcommand{\numrepeats}{n_\text{repeat}}
\newcommand{\terminate}{\epsilon}
\newcommand{\norm}{\mathcal{N}}
\newcommand{\polystan}{\pkg{PolyStan}}

\input{jssifarxiv.tex}
\arxivtrue

\author{Andrew Fowlie~\orcidlink{0000-0003-0918-3766}\\ Xi'an Jiaotong-Liverpool University}

\Plainauthor{Andrew Fowlie}

\title{\polystan{}: \pkg{PolyChord} nested sampling and Bayesian evidences for \proglang{Stan} models}

\Plaintitle{PolyStan: PolyChord nested sampling and Bayesian evidences for Stan models}

\Shorttitle{Nested sampling for \proglang{Stan} models}

\Abstract{\noindent
  Sampling from multi-modal distributions and estimating marginal likelihoods, also known as evidences and normalizing constants, are well-known challenges in statistical computation. They can be overcome by nested sampling, which evolves a set of live points through a sequence of distributions upwards in likelihood. We introduce \polystan{} --- a nested sampling inference engine for \proglang{Stan}. \polystan{} provides a \proglang{Stan} interface to the \pkg{PolyChord} nested sampling algorithm using \pkg{bridgestan}.  \polystan{} introduces a new user-base to nested sampling algorithms and provides a black-box method for sampling from challenging distributions and computing marginal likelihoods. We demonstrate the robustness of nested sampling on several degenerate and multi-modal problems, comparing it to bridge sampling and Hamiltonian Monte Carlo.
}
\Keywords{nested sampling, marginal likelihood, model selection, Bayes factor, multi-modal, \proglang{Stan}}
\Plainkeywords{nested sampling, marginal likelihood, model selection, Bayes factor, multi-modal, Stan}

\Address{
  Andrew Fowlie\\
  X-HEP Laboratory\\
  Department of Physics\\
  School of Mathematics and Physics\\ Xi'an Jiaotong-Liverpool University\\ Suzhou, 215123, China\\
  E-mail: \email{Andrew.Fowlie@XJTLU.edu.cn}\\
  URL: \url{https://andrewfowlie.github.io/}
}

\begin{document}

\section{Introduction}

\proglang{Stan}~\citep{stan} is a probabilistic programming language for statistical modeling and a suite of state-of-the-art algorithms for statistical computation. The built-in algorithms enable parameter estimation e.g.~through the posterior by Hamiltonian Monte Carlo (HMC)~\citep{Neal:2011mrf,Betancourt:2017ebh} or maximum likelihood estimation. On the other hand, one may wish to compute model evidences and perform model selection through Bayes factors~\citep[see e.g.][]{Jeffreys:1939xee,Kass01061995}. 

The evidence $\ev$ plays an important role in Bayesian model selection, as the ratio of evidences for two models,
\begin{equation}
    B_{10} = \frac{\ev_1}{\ev_0},
\end{equation}
tells us the change in the relative plausibly of those two models,
\begin{equation}
    \text{Posterior odds} = \text{Bayes factor} \times \text{Prior odds}.
\end{equation}
The evidence for a model with parameters $\params$ may be found from the integral,
\begin{equation}\label{eq:ev}
    \ev \equiv \int \like(\params) \, \pi(\params) \dint \params 
\end{equation}
for a likelihood function $\like$ and a choice of prior for the model's parameters, $\pi$. The integral \cref{eq:ev}, however, can be challenging to compute, especially in high dimensions. 
Bayes factors can be estimated from the posterior through, e.g., Savage-Dickey ratios~\citep{10.1214/aoms/1177693507} or bridge sampling~\citep{meng1996simulating}. The latter is implemented in the \pkg{bridgesampling} package in \proglang{R}~\citep{Gronau2020}.

Here, we introduce a \proglang{Stan} interface to the nested sampling algorithm~\citep[NS;][]{Skilling2004,10.1214/06-BA127}. NS is a powerful black-box algorithm for tackling challenging integrals. As a byproduct, it produces posterior samples at the same time~\citep[see e.g.~the review][]{Ashton:2022grj}. NS copes with challenging posterior distributions, including discrete parameters annd multi-modal and degenerate distributions, which can be challenging in HMC~\citep{park2021sampling}. 
We try to match \pkg{cmdstan} interfaces as closely as possible for both building and running programs; thus, users familiar with \pkg{cmdstan} should be able to easily build and run programs.

\subsection{Nested Sampling}

In nested sampling (NS), we rewrite the integral \cref{eq:ev} as
\begin{equation}
    \ev \equiv \int_0^1 \like(X)  \dint X,
\end{equation}
where the volume variable,
\begin{equation}
    X(\like^\star) = \int\limits_{\like \ge \like^\star} \pi(\params) \dint \params. 
\end{equation}
This turns our multidimensional integral \cref{eq:ev} into a one-dimensional integral. We proceed by making statistical estimates of $X$ at contours of the likelihood, enabling us to approximate the evidence integral by
\begin{equation}\label{eq:ev_sum}
    \ev \simeq \sum_i \like_i \, \Delta X_i.
\end{equation}
We now sketch the procedure by which the statistical estimates of the volume are obtained. We start by drawing $\nlive$ points from the prior --- these points are uniform draws from the entire volume $X = 1$. We denote the minimum likelihood among these points by $\like^\star$. We evict it from the live points and sample a replacement from the constrained prior
\begin{equation}\label{eq:constrained_prior}
    \pi^\star(\params) \propto \begin{cases}
        \pi(\params) & \text{if}~\like(\params) > \like^\star\\
        0 & \text{otherwise}
    \end{cases}
\end{equation}
Since the $\nlive$ live points were distributed uniformly in $X$ from $0$ to $1$, we estimate that the volume compressed by a factor
\begin{equation}
\frac{\nlive}{\nlive + 1}
\end{equation}
We repeat this process, replacing the worst live point and compressing upwards in likelihood. At iteration $i$, we estimate the volume by
\begin{equation}\label{eq:volume_estimate}
    X_i = \left(\frac{\nlive}{\nlive + 1}\right)^i \approx e^{-i/\nlive}
\end{equation}
and increment the evidence by \cref{eq:ev_sum}. We stop once the sum \cref{eq:ev_sum} has appeared to converge.  To assess convergence, we estimate the remaining unsummed evidence by the average likelihood of the current live points multiplied by the current volume estimate,
\begin{equation}
   \Delta \ev = \langle \like \rangle X.
\end{equation}
For convergence, we require that the relative impact of this remaining evidence is less than a user-specified tolerance,
\begin{equation}
    \frac{\Delta \ev}{\ev} < \epsilon.
\end{equation}
Of course, \cref{eq:volume_estimate} represents only a statistical estimate of the volume from the NS procedure. For more refined estimators and an analysis of the uncertainty in the NS method, the distribution of the compression factors can be analyzed.

\subsection{The PolyChord algorithm}\label{sec:polychord}

The challenging aspect of an efficient and robust NS implementation is designing a scheme to draw from the constrained prior in \cref{eq:constrained_prior}, as rejection sampling from the prior could be extremely inefficient. Generally, we anticipate that the the cost of rejection sampling schemes scales exponentially with dimension. Metropolis-Hastings schemes were originally proposed~\citep{Skilling2004,10.1214/06-BA127} as a solution and could scale polynomially, but the proposals could require problem-specific fine-tuning. 

\cite{Neal2003} proposed slice sampling as a way to sample from a posterior distribution by randomly sampling from slices $P(\params) \ge P^\star$; this is reminiscent of the problem of sampling from the constrained prior. \cite{Handley:2015fda,Handley:2015vkr} \citep[see also][]{Aitken2013} thus proposed slice sampling as a solution. We now briefly review the \pkg{PolyChord} algorithm developed and implemented by~\cite{Handley:2015fda,Handley:2015vkr}. 

In the \pkg{PolyChord} algorithm, one starts a Markov chain from a randomly chosen live point. The next point is found by slice sampling from $\like > \like^\star$ along a randomly chosen direction. The Markov chain is extended by repeating this $\numrepeats$ times. If enough repeats were performed, the final point should be de-correlated from the initial live point and may be used as an independent draw from the constrained prior. There are additional whitening and clustering steps to deal with multi-modal and degenerate posterior distributions.

There are three main parameters that control the \pkg{PolyChord} algorithm:
\begin{itemize}
    \item $\nlive$ --- The number of live points. The effective sample size and runtime scale as $\mathcal{O}(\nlive)$; the error on the evidence estimate scales as $\mathcal{O}(1/\sqrt{\nlive}\,)$.
    \item $\numrepeats$ --- The number of slice sampling steps. The runtime scales as $\mathcal{O}(\numrepeats)$; the accuracy of the results depends on this parameter.
    \item $\terminate$ --- The stopping condition --- we stop once $\Delta \ev / \ev < \terminate$.  The runtime and truncation error depend on this parameter.
\end{itemize}
While they are set to reasonable defaults, in some cases they may need to tuned per problem.

\subsection{Insertion index test}\label{sec:test}

Problematic HMC runs can be identified by e.g.~divergent transitions at runtime or by computing diagnostics on the MCMC chains. For example, the $\hat R$ parameter~\citep{Vehtari_2021} tests whether inter and intra-chain variances are consistent with convergence. In a similar vein, a so-called insertion index test may be performed on NS runs~\citep{Fowlie:2020mzs}. 

The critical question in an NS run is whether samples were faithfully obtained from the constrained prior. If they were, when ranked by likelihood, the insertion index of new live points into existing live points should be uniformly distributed from $0$ to $\nlive$. If, on the other hand, the NS implementation draws new live points from different distribution, the new live points will not be uniformly inserted into the existing ones. We can thus test whether the insertion indexes observed in an NS run are consistent with a uniform distribution. Similarly to~\cite{Vehtari_2021}, this uses ranking and is thus invariant to reparameterisation. 

In~\cite{Fowlie:2020mzs} and in \polystan{}, we report a $p$-value from a Kolmogorov-Smirnov test~\citep{Massey1951}. This is straight-forward though may not be the optimal choice~\citep{Buchner:2021kpm}. Small $p$-values indicate that the NS results may be unreliable.

\section{Quickstart}

\subsection{Obtaining code}

\polystan{} requires some basic development tools to build \pkg{bridgestan} and \pkg{PolyChord}:
\begin{CodeChunk}
\begin{CodeInput}[bash]
sudo apt-get install git make gcc gfortran  # on Ubuntu/debian
sudo dnf install git make gcc gfortran  # on Fedora
\end{CodeInput}
\end{CodeChunk}
If you wish speed-up computations using multiple processors using \pkg{Open MPI} you must install 
\begin{CodeChunk}
\begin{CodeInput}[bash]
sudo apt-get install libopenmpi-dev # on Ubuntu/debian
sudo dnf install openmpi openmpi-devel  # on Fedora
module load mpi/openmpi-$(uname -m)  # on Fedora
\end{CodeInput}
\end{CodeChunk}
We must obtain the code through
\begin{CodeChunk}
\begin{CodeInput}[bash]
git clone --recursive https://github.com/xhep-lab/polystan
\end{CodeInput}
\end{CodeChunk}
This must be recursive to include \pkg{bridgestan}~\citep{Roualdes2023} and \pkg{PolyChord}~\citep{Handley:2015fda,Handley:2015vkr} .

\subsection{Running toy example}\label{sec:toy}

As an example, we consider a standard example in \proglang{Stan} --- the Bernoulli model. The model is specified in \code{examples/bernoulli.stan} as,
\ifarxiv%
\begin{lstlisting}[language=stan, mathescape, numbers=left, title=\texttt{polystan/examples/bernoulli.stan}]
functions {
  #include polystan.stanfunctions
}
data {
  int<lower=0> N;
  array[N] int<lower=0, upper=1> y;
}
parameters {
  real<lower=0, upper=1> x;
}
transformed parameters {
  real theta = beta_prior(x, 1, 1);
  real logit_theta = logit(theta);
}
model {
  y ~ bernoulli(theta);
}
generated quantities {
  int y_sim = bernoulli_rng(theta);
}
\end{lstlisting}
\else%
\begin{CodeChunk}
\verbatiminput{bernoulli.stan}
\end{CodeChunk}
\fi%
As usual, \proglang{Stan} model files must end with extension \code{.stan}.
This \proglang{Stan} model file must be compiled by \pkg{make} from within the \polystan{} directory. E.g.,
\begin{CodeChunk}
\begin{CodeInput}[bash]
make examples/bernoulli
\end{CodeInput}
\end{CodeChunk}
Like the \pkg{cmdstan} interface, we specify the desired executable as a \pkg{make} target. This should be the name of the \proglang{Stan} model file without the \code{.stan} extension. Intermediate output files, such as the transpiled \proglang{C++} header file, are stored in the \code{build} directory.

Just like \pkg{cmdstan} models, to run the model, we may need to specify a data file to set entries in the \code[stan]{data} block. In this case, we provide \code[none]{bernoulli.data.json},
\begin{CodeChunk}
\begin{CodeInput}[bash]
cd examples
./bernoulli data --file=bernoulli.data.json
\end{CodeInput}
\end{CodeChunk}
This runs the \pkg{PolyChord} algorithm on our model with default settings. This differs slightly from \pkg{cmdstan} where the command would be \code{data file=bernoulli.data.json}; we implement the CLI by  \pkg{CLI11}~\citep{cli11} which distinguishes subcommands from options.

By default, models are compiled with \pkg{Open MPI} if it is installed. To run with \pkg{Open MPI}, try e.g.,
\begin{CodeChunk}
\begin{CodeInput}[bash]
mpirun -n 4 bernoulli data --file=bernoulli.data.json
\end{CodeInput}
\end{CodeChunk}
for 4 cores. The main result of the run should be printed to the screen at the end, e.g.,
 \begin{CodeChunk}
\begin{CodeOutput}[pso]
 ____________________________________________________ 
|                                                    |
| ndead  =         4562                              |
| log(Z) =           -6.21093 +/-            0.04053 |
|____________________________________________________|

| Finished PolyChord
| Native PolyChord results at /home/andrew/repos/polystan/chains/bernoulli*
| PolyStan JSON summary at /home/andrew/repos/polystan/bernoulli.json
| 
| Evidence log(Z) = -6.21093 +/- 0.0405311
| P-value of sampling from constrained prior = 0.687816
| Effective number of samples = 1022
| 
| If you use these results, you are required to cite
| https://arxiv.org/abs/1502.01856
| https://arxiv.org/abs/1506.00171
| and agree to the license
| https://github.com/PolyChord/PolyChordLite/raw/refs/heads/master/LICENCE
\end{CodeOutput}
\end{CodeChunk}
We see the estimate $\log Z = -6.20258 \pm 0.0408675$.

\section{Program options}

\subsection{Compilation options}

\newcommand{\buildvar}[3]{\code{#1 [ALLOWED=#2. DEFAULT=#3]}}

\polystan{} builds and links to \pkg{PolyChord}  and \pkg{bridgestan}. To build and link consistently with the \pkg{PolyChord} library, a build variable \buildvar{MPI}{0, 1}{1 if MPI installed else 0} controls whether \pkg{Open MPI} is enabled. E.g.,
\begin{CodeChunk}
\begin{CodeInput}[bash]
make example/bernoulli MPI=0
\end{CodeInput}
\end{CodeChunk}
disables \pkg{Open MPI}. To enable debugging, a build variable  \buildvar{DEBUG}{0, 1}{1} disables optimizations and enables debugging flags. E.g.,
\begin{CodeChunk}
\begin{CodeInput}[bash]
make example/bernoulli DEBUG=1
\end{CodeInput}
\end{CodeChunk}
Lastly, standard build variables can control compiler choices, e.g., \buildvar{CXX}{g++, clang++, etc}{g++}.

\subsection{Runtime options}

The runtime options can be found, by e.g.,
\begin{CodeChunk}
\begin{CodeInput}[bash]
./bernoulli --help
\end{CodeInput}
\end{CodeChunk}
The available options consist of flags and subcommands.

\subsubsection{Flags}

There are \code{--version}, \code[none]{--polychord-version}, and \code{--stan-file-name} flags to show program version numbers and the name of the original \proglang{Stan} file.

There is, furthermore, a \code{--from-toml} flag that enables settings to be specified in a \code{toml} file. As these \code{toml} files are written automatically by \polystan{}, this flag enables runs to be completely repeated with explicit inputs.

\subsubsection{Subcommands}

There are \code{polychord}, \code{random}, \code{data} and \code{output} subcommand options, similar to \pkg{cmdstan}. See e.g.,
\begin{CodeChunk}
\begin{CodeInput}[bash]
./bernoulli polychord --help 
\end{CodeInput}
\end{CodeChunk}
for help about the \code{polychord} subcommand. These subcommands enable one to control \pkg{PolyChord} algorithm parameters, random number generation in \proglang{Stan}, the \code[stan]{data} block and \code{polystan} output files, respectively. E.g., to change the program's  data file name and the \code{json} output file name, 
\begin{CodeChunk}
\begin{CodeInput}[bash]
./bernoulli data --file=bernoulli.data.json output --json-file=this_file_name.json
\end{CodeInput}
\end{CodeChunk}
\proglang{Stan} and \pkg{PolyChord} random number generation are controlled separately. To set them both,
\begin{CodeChunk}
\begin{CodeInput}[bash]
./bernoulli data --file=bernoulli.data.json random --seed=10 polychord --seed=20
\end{CodeInput}
\end{CodeChunk}
The \proglang{Stan} random number generation concerns only the \code[stan]{generated quantities} block. 

For the \pkg{PolyChord} subcommand, the three important algorithm parameters described in \cref{sec:polychord}, $\nlive$, $\numrepeats$ and $\terminate$, can be set by e.g.,
\begin{CodeChunk}
\begin{CodeInput}[bash]
./bernoulli data --file=bernoulli.data.json polychord --nlive 1000 --num-repeats 5 --precision 0.0001
\end{CodeInput}
\end{CodeChunk}
They default to 500, 5 repeats per dimension, and 0.001, respectively. There are three special flags,
\begin{CodeChunk}
\begin{CodeInput}[bash]
./bernoulli polychord --no-feedback --no-write --no-derived
\end{CodeInput}
\end{CodeChunk}
that do not correspond to standard \pkg{PolyChord} options. They turn off feedback that is printed to the screen, turn off all writing to disk, and turn-off storing derived parameters in memory or on disk, respectively. These are helpful if one is only interested in the evidence estimate and for fast runs in which writing to disk is a bottleneck. 

In particular \code{--no-derived} was provided to work around a limitation of the \proglang{Stan} interface. To obtain derived parameters, e.g.~the \code[stan]{transformed parameters} or \code[stan]{generated parameters} block, and the target function from \proglang{Stan},  the derived parameters must be computed twice: once explicitly and once implicitly in the computation of the target function. If they are expensive to compute, this is wasteful. The\code{--no-derived} flag ensures they are only computed once, though as a result they cannot be saved to disk.

\section{Program outputs}

\subsection{Printed to screen}

The blue output at the beginning and end of a run originates from \polystan{} itself. This shows e.g., program version information, where results are saved to disk, estimates of the evidence and the effective sample size, and a $p$-value from a test of sampling from the constrained prior. The other output originates from \pkg{PolyChord} and shows the progress and final result of the NS algorithm.

\subsection{Saved to disk}

As well as the results printed to the screen by \pkg{PolyChord}, \pkg{PolyChord} saves results to text files named after the model e.g., \code{chains/bernoulli*}. These native \pkg{PolyChord} files allow runs to be resumed, are described in the \pkg{PolyChord} documentation.

In addition, \polystan{} saves a summary of the results to a \code{JSON} file named after the model e.g., \code{bernoulli.json}. The \polystan{} summary contains metadata about the run and the main results in \code{JSON} format. E.g., for our toy problem Bernoulli model, the metadata entries of \code{bernoulli.json} include:
 \begin{CodeChunk}
\begin{CodeOutput}[json]
    "posterior_attrs": {
        "name": "bernoulli_model",
        "created_at": "Thu Apr 24 14:19:02 2025",
        "inference_library": "PolyChord",
        "inference_library_version": "1.22.1",
        "creation_library": "PolyStan",
        "creation_library_version": "1.0.0",
        "creation_library_language": "C++",
        "polystan": {
            "stan file name": "/polystan/examples/bernoulli.stan",
            "stan data file": "/polystan/examples/bernoulli.data.json",
            "polystan toml file": "/polystan/bernoulli.toml",
            "stan build info": "BridgeStan version: 2.6.1\nStan version: 2.35.0\nStan C++ Defines:\nSTAN_THREADS=false\nSTAN_MPI=false\nSTAN_OPENCL=false\nSTAN_NO_RANGE_CHECKS=true\nSTAN_CPP_OPTIMS=false\nBRIDGESTAN_AD_HESSIAN=false\nStan Compiler Details:\nstanc_version = stanc3 v2.35.0\nstancflags = --include-paths /polystan/stanfunctions\n",
            "seed": 0
        },
        "polychord": {
            "nlive": 500,
            "num_repeats": 5,
            "precision_criterion": 0.001,
            "seed": -1
        }
    },
\end{CodeOutput}
\end{CodeChunk}
They record version numbers and build information, and basic run settings. The complete run settings are recorded in \code{bernoulli.toml}.

The data entries include sample statistics,
\begin{CodeChunk}
\begin{CodeOutput}[json]
    "sample_stats": {
        "test": {
            "metadata": "This is a test of uniformity of insertion indexes of live points",
            "p-value": 0.36257046605026457,
            "batch size / n_live": 1
        },
        "ess": {
            "metadata": "Estimate of effective sample size",
            "n": 979
        },
        "evidence": {
            "metadata": "The evidence is log-normally distributed",
            "log evidence": -6.202493190765381,
            "error log evidence": 0.040782634168863299
        },
        "neval": {
            "metadata": "Total number of likelihood evaluations",
            "neval": 86408
        }
    }
\end{CodeOutput}
\end{CodeChunk}
These show our evidence estimate, effective sample size, and $p$-value. Lastly, there are entries \code{posterior} and \code{prior} that contain equally-weighted samples from the posteror and prior, respectively. These include the target \code{log_likelihood}, and parameters in the \code[stan]{parameters}, \code[stan]{transformed parameters} and \code[stan]{generated quantities} blocks.

\subsection{Analyzing results}

\begin{figure}
    \centering
    \includegraphics[width=0.5\linewidth]{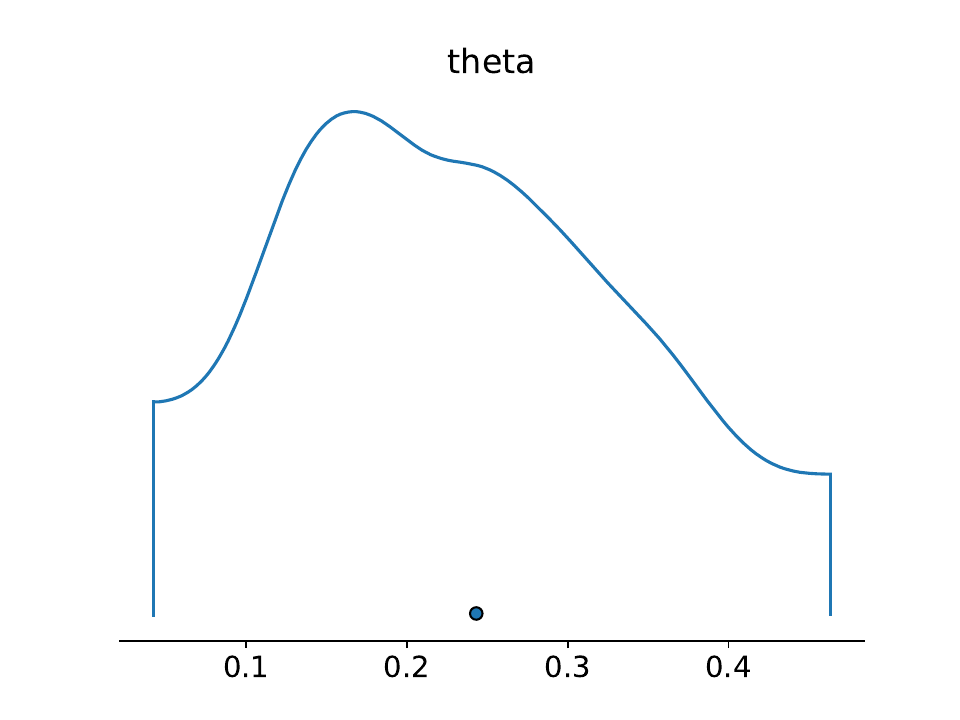}
    \caption{Posterior pdf for \code{theta} in the \code{bernoulli.stan} model, found by \polystan{} and plotted with \pkg{ArviZ}~\citep{arviz_2019} and  \pkg{matplotlib}~\citep{Hunter:2007}.}
    \label{fig:bernoulli}
\end{figure}

The results can be readily loaded, e.g., in \proglang{Python} using \pkg{ArviZ}~\citep{arviz_2019} we can inspect the results from the Bernoulli model,
\begin{CodeChunk}
\begin{CodeInput}[Python3Prompt]
import arviz as az
data = az.from_json("bernoulli.json")
az.plot_density(data, var_names="theta", show=True)
\end{CodeInput}
\end{CodeChunk}
This shows \cref{fig:bernoulli} --- the posterior pdf for the parameter \code{theta}. The package \pkg{ArviZ} requires installation. This could be achieved by e.g., \code[bash]{pip install arviz}.

You may wish to inspect the priors produced by the inverse transform method. As prior samples are saved by default, this can be done by
\begin{CodeChunk}
\begin{CodeInput}[Python3Prompt]
az.plot_density(data, group="prior", var_names="theta", show=True)
\end{CodeInput}
\end{CodeChunk}

Lastly, the native \pkg{PolyChord} outputs can be further analyzed with \pkg{anesthetic}~\citep{anesthetic}, which can be installed by e.g. \code[bash]{pip install anesthetic}. We load the data, including adding labels, using
\begin{CodeChunk}
\ifarxiv%
\begin{CodeInput}[Python3]
@\pyprompt{}@ from anesthetic import read_chains
@\pyprompt{}@ nested_samples = read_chains("chains/bernoulli", labels={
@\pycontin{}@                              0: "$x$",
@\pycontin{}@                              1: r"$\theta$",
@\pycontin{}@                              2: r"$\text{logit}\,\theta$",
@\pycontin{}@                              3: r"$y_{\text{sim}}$"})
\end{CodeInput}
\else%
\begin{CodeInput}[Python3]
from anesthetic import read_chains
nested_samples = read_chains("chains/bernoulli", labels={
                             0: "$x$",
                             1: r"$\theta$",
                             2: r"$\text{logit}\,\theta$",
                             3: r"$y_{\text{sim}}$"})
\end{CodeInput}
\fi%
\end{CodeChunk}
We can plot posterior and prior distributions, e.g.,
\begin{CodeChunk}
\begin{CodeInput}[Python3Prompt]
import matplotlib.pyplot as plt
nested_samples.plot_1d()
plt.show()
\end{CodeInput}
\end{CodeChunk}
We can, furthermore, recompute the evidence and other statistics connected to the evidence integral through
\begin{CodeChunk}
\begin{CodeInput}[Python3Prompt]
nested_samples.stats()
\end{CodeInput}
\begin{CodeOutput}[none]
        labels                                    
logZ    $\ln\mathcal{Z}$                             -6.262698
D_KL    $\mathcal{D}_\mathrm{KL}$                     0.789167
logL_P  $\langle\ln\mathcal{L}\rangle_\mathcal{P}$   -5.473532
d_G     $d_\mathrm{G}$                                0.859294
dtype: float64
\end{CodeOutput}
\end{CodeChunk}
We can inspect the run interactively using a graphical user interface,
\begin{CodeChunk}
\begin{CodeOutput}[Python3Prompt]
nested_samples.gui()
\end{CodeOutput}
\end{CodeChunk}
Lastly, we can compute the insertion index test in \cref{sec:test},
\begin{CodeChunk}
\ifarxiv%
\begin{CodeInput}[Python3]
@\pyprompt{}@ from anesthetic.utils import compute_insertion_indexes, insertion_p_value
@\pyprompt{}@ indexes = compute_insertion_indexes(
@\pycontin{}@     nested_samples.logL.to_numpy(), nested_samples.logL_birth.to_numpy())
@\pyprompt{}@ nlive = 500 # our PolyStan run used @\color{comment}500@ live points
@\pyprompt{}@ insertion_p_value(indexes, nlive)
\end{CodeInput}
\else%
\begin{CodeInput}[Python3]
from anesthetic.utils import compute_insertion_indexes, insertion_p_value
indexes = compute_insertion_indexes(
    nested_samples.logL.to_numpy(), nested_samples.logL_birth.to_numpy())
nlive = 500 # our PolyStan run used 500 live points
insertion_p_value(indexes, nlive)
\end{CodeInput}
\fi%
\begin{CodeOutput}[Python3]
{'D': np.float64(0.010760244115083029), 'sample_size': 4588, 'p-value': np.float64(0.6628378526651819)}
\end{CodeOutput}
\end{CodeChunk}

\section{Supported models}

\polystan{} models are entirely valid \proglang{Stan} models --- we deliberately do not introduce any new syntax or any new keywords. A user must, however, take care to make sure that the model distinguishes the prior and the likelihood, as discussed below.

\subsection{Specifying parameters}

Whereas in MCMC algorithms the prior and likelihood appear always as a product, NS distinguishes the prior from the likelihood. To enable that distinction in a \proglang{Stan} model, 
\begin{itemize}   
    \item The \code[stan]{parameters} block should declare parameters on a unit hypercube, that is, parameters must be constrained to between 0 and 1, e.g., 
\begin{CodeChunk}
\begin{CodeInput}[Stan]
real <lower=0, upper=1> x;
\end{CodeInput}
\end{CodeChunk}
    \emph{This is enforced at runtime --- parameters that are not constrained to $[0, 1]$ result in a fatal exception.}
    
\item The \code[stan]{transformed parameters} block should transform the unit hypercube parameters into the physical parameters using inverse transform sampling, e.g.,
\begin{CodeChunk}
\begin{CodeInput}[Stan, mathescape]
real theta = inv_Phi(x);  // $\theta \sim \mathcal{N}(0, 1)$
\end{CodeInput}
\end{CodeChunk}
    \emph{This cannot be enforced --- there is no way of knowing the user's intent.}
    
    \item The \code[stan]{model} block should only increment the target with contributions to the likelihood.

    \emph{This cannot be enforced --- there is no way of knowing the user's intent.}
\end{itemize}
For example, suppose we wish to have a parameter $\theta$ with prior,
\begin{equation}
    \theta \sim \mathcal{N}(\mu, \sigma^2)
\end{equation}
We achieve this through inverse transform sampling from the unit hypercube as follows. First, we draw from the unit hypercube,
\begin{equation}
    x \sim \mathcal{U}(0, 1)
\end{equation}
Then we transform that draw using the inverse cumulative distribution for the normal distribution,
\begin{equation}
\theta = \mu + \sigma \, \Phi^{-1}(x).
\end{equation}

In a \proglang{Stan} model, we could define the mean and variance in the \code[stan]{data} block,
\begin{CodeChunk}
\begin{CodeInput}[Stan, mathescape]
data {
  real mu; // $\mu$
  real sigma; // $\sigma$
}
\end{CodeInput}
\end{CodeChunk}
We would define the unit hypercube parameter in the \code[stan]{parameters} block,
\begin{CodeChunk}
\begin{CodeInput}[Stan, mathescape]
parameters {
  real<lower=0, upper=1> x; // $x \sim \mathcal{U}(0,1)$
}
\end{CodeInput}
\end{CodeChunk}
and transform that parameter in the \code[stan]{transformed parameters} block,
\begin{CodeChunk}
\begin{CodeInput}[Stan, mathescape]
transformed parameters {
  real theta = mu + sigma * inv_Phi(x); // $\theta = \mu + \sigma \, \Phi^{-1}(x)$
}
\end{CodeInput}
\end{CodeChunk}
We would not include any contribution to the target function from the prior in the \code[stan]{model} block.

This procedure is completely general, though requires knowledge of the inverse cumulative distribution function. For convenience, we provide a \code[stan]{polystan.stanfunctions} header containing transforms for common choices of prior.  See \cref{app:prior_transforms} for details. To use them, 
\ifarxiv%
\code[stan]{#include polystan.stanfunctions} 
\else%
\code[stan]{\#include polystan.stanfunctions} 
\fi%
in the \code[stan]{functions} block.

\subsection{Normalization}

Lastly, the normalization of the likelihood changes the evidence estimate by a constant factor, $\like \to C \like$ implies $\ev \to C \ev$, though leaves the posterior unchanged. Since we are concerned with the evidence, in \polystan{} we retain normalization constants. That is, distribution statements, and \code[stan]{lupdf} and \code[stan]{lupmf} functions, e.g.,
\begin{CodeChunk}
\begin{CodeInput}[Stan]
model {
  x ~ std_normal();  // distribution statement
  target += std_normal_lupdf(x);  // lupdf function
}
\end{CodeInput}
\end{CodeChunk}
increment the target function by a \emph{normalized} contribution. This could be somewhat surprising, as in \pkg{cmdstan} a constant $\log\sqrt{2\pi}$ term would be discarded. \polystan{} accesses the target through \pkg{BridgeStan}, specifying \code[cpp]{propto=false} to disable this behavior. See \citet[\href{https://mc-stan.org/docs/reference-manual/statements.html\#log-probability-increment-vs.-distribution-statement}{Log probability increment vs.\ distribution statement}]{stan} for further discussion. 

\section{Examples}

The \polystan{} source code comes with several example programs in the \code{polystan/examples} directory. Many of these examples are \proglang{Stan} implementations of problems supplied in \code{Fortran} with \pkg{PolyChord} to showcase performance in high-dimensional, multi-modal and degenerate problems.

We now discuss some of these examples, showing results from both \polystan{} and \pkg{bridgesampling} using \proglang{Stan} HMC chains.
Although we do not make a detailed comparison between HMC, \polystan{} and \pkg{bridgesampling}, our examples illustrate that \polystan{} is a useful addition to the toolbox. We used \polystan{} and \pkg{bridgesampling} with default settings, and HMC with 20,000 iterations, 5,000 warmup iterations and 4 chains. The number of target evaluations of the algorithms should be taken with a grain of salt --- they depend on tuning parameters. 

As emphasized in \cite{Gronau2020}, the quality of \pkg{bridgesampling} marginal likelihood estimates depends on the quality of the posterior samples. Thus, as we shall see, when HMC fails to converge,  \pkg{bridgesampling} produces faulty marginal likelihood estimates.

\newcommand{\twosubexample}[2]{\subsection*{#1 --- \texttt{polystan/examples/#2.\{stan,data.json\}}}}
\newcommand{\subexample}[1]{\subsection*{\MakeUppercase #1 --- \texttt{polystan/examples/#1.\{stan,data.json\}}}}

\subexample{bernoulli}

This is the classic, one-dimensional problem in \cref{sec:toy}. The evidence estimates are consistent,
\begin{align}
\text{\polystan{}:}~\log\ev &= -6.20 \pm 0.04~\text{[0.1 million evaluations]}\\
\text{\pkg{bridgestan}:}~\log\ev &= -6.2048 \pm 0.0004~\text{[0.2 million evaluations]}
\end{align}
as expected from a simple uni-modal problem. There was a warning that the ESS could not be computed and the number of samples would be used instead. Thus, the \pkg{bridgesampling} error could be an underestimate.

\subexample{eggbox}

This is a challenging multi-modal problem in $N = 10$ dimensions. The target function,
\begin{equation}
    \log\like = -\left(2 + \prod_{i=1}^{N}  \cos(\theta_i / 2)\right)^5
\end{equation}
resembles an eggbox with regularly spaced modes. HMC sampling and thus \pkg{bridgesampling} failed, producing faulty evidence estimates:
\begin{align}
\text{\polystan{}:}~\log\ev &= -14.9 \pm 0.1~\text{[1.9 million evaluations]}\\
\text{\pkg{bridgestan}:}~\log\ev &= -28.3 \pm 0.1~\text{[0.8 million evaluations]}
\end{align}
We found similar estimates with \code{method=warp3}.
Failure was indicated by warnings about $\hat R \gg 1$, the ESS and divergent transitions. 

\subexample{himmelblau}

The two-dimensional Himmelblau function,
\begin{equation}
    \log\like = \text{const.} - (\theta_1^2 + \theta_2 - 11)^2 - (\theta_1 + \theta_2^2 - 7)^2
\end{equation}
contains four dengenerate modes. The results,
\begin{align}
\text{\polystan{}:}~\log\ev &= -4.66 \pm 0.1~\text{[0.5 million evaluations]}\\
\text{\pkg{bridgestan}:}~\log\ev &= -4.85 \pm 0.03~\text{[0.3 million evaluations]}
\end{align}
were consistent, despite warnings about ESS and $\hat R \gg 1$ from the HMC chains.

\subexample{rastrigin}

This is a complicated multi-modal function in $N = 10$ dimensions,
\begin{equation}
    \log\like = \text{const.} - \sum_{i=1}^N \theta_i^2 - 10 \cos(2 \pi \theta_i)
\end{equation}
We found faulty estimates from HMC and \pkg{bridgesampling},
\begin{align}
\text{\polystan{}:}~\log\ev &= -23.4 \pm 0.2~\text{[6.1 million evaluations]}\\
\text{\pkg{bridgestan}:}~\log\ev &= -104.21 \pm 0.05~\text{[0.6 million evaluations]}
\end{align}
and similar estimates with \code{method=warp3}. Failure was indicated by warnings about ESS and $\hat R \gg 1$.

\subexample{rosenbrock}

This is an $N=4$ dimensional problem defined by
\begin{equation}
    \log\like = \text{const.} - \sum_{i=1}^{N-1} 100 (x_{i+1} - x_i^2)^2 - (1 - x_i)^2
\end{equation}
In $N = 4$ there are two modes. The estimates are consistent,
\begin{align}
\text{\polystan{}:}~\log\ev &= -9.5 \pm 0.2~\text{[1.1 million evaluations]}\\
\text{\pkg{bridgestan}:}~\log\ev &= -9.39 \pm 0.02~\text{[2.2 million evaluations]}
\end{align}
despite the multi-modality in this problem. There were no warnings about ESS or $\hat R$.

\subexample{shell}

This is a Gaussian shell,
\begin{equation}
    \log\like = - \frac{(r^2 - \mu)^2}{2\sigma^2}
\end{equation}
for $\mu = 0.25$, $\sigma = 0.01$ and radius $r^2 = \sum_{i=1}^N x_i^2$ for $N  = 5$. The evidence estimates were consistent,
\begin{align}
\text{\polystan{}:}~\log\ev &= -5.8 \pm 0.1~\text{[0.9 million evaluations]}\\
\text{\pkg{bridgestan}:}~\log\ev &= -5.71 \pm 0.01~\text{[3.9 million evaluations]}
\end{align}
as HMC navigated the geometry of the shell.

\twosubexample{Slab \& spike}{slab\_spike}

This a one-dimensional Gaussian spike on top of a Gaussian slab,
\begin{equation}
    \like = \norm\!(x \,|\, 0, \sigma_1^2) + 
            \norm\!(x \,|\, 0, \sigma_2^2)
\end{equation}
for $\sigma_1 = 50$ and $\sigma_2 = 0.01$. The hierarchy of scales $\sigma_1 \gg \sigma_2$ was challenging for HMC and resulted in disagreement,
\begin{align}
\text{\polystan{}:}~\log\ev &= -4.63 \pm 0.07~\text{[0.2 million evaluations]}\\
\text{\pkg{bridgestan}:}~\log\ev &= -5.3407 \pm 0.0007~\text{[0.3 million evaluations]}
\end{align}
This was indicated by warnings about divergent transitions.

\twosubexample{GLMM}{glmm\_\{h0, h1\}}

These are generalized linear mixed models (GLMM) for the turtles dataset~\citep{Janzen2000}, presented as examples in \cite{Gronau2020}. The data consists of the birth weights $m$, survival status $s$, and clutch membership $c$ of $n = 244$ turtles from 31 clutches. The null hypothesis models the survival of turtles using their birth weights. The survival probability for turtle~$i$,
\begin{equation}
    p_i = \Phi(\alpha_0 + \alpha_1 m_i),
\end{equation}
where $\alpha_0$ and $\alpha_1$ are two unknown parameters and $\Phi$ is the standard normal cumulative density function. There are normal priors for the  $\alpha_0$ and $\alpha_1$ parameters.

The alternative model extends the null by a clutch-dependent effect on survival. The survival probability for turtle $i$ becomes
\begin{equation}
    p_i = \Phi(\alpha_0 + \alpha_1 m_i + \beta_{c_i}),
\end{equation}
where $c_i$ is the clutch of turtle $i$ and $\beta_c$ is the clutch effect parameter for clutch $c$. There are independent and identical normal priors for the 33 $\beta$ parameters. 

In each model, the likelihood is a product of Bernoulli pmfs,
\begin{equation}
    \like = \prod_{i=1}^n \text{Bern}(s_i \,|\, p_i),
\end{equation}
for the survival of each turtle, where $s_i$ indicates whether the turtle survived. 

There was agreement between \polystan{} and  \pkg{bridgestan} for the evidence for the null,
\begin{align}
\text{\polystan{}:}~\log\ev_0 &= -156.3 \pm 0.1~\text{[0.3 million evaluations]}\\
\text{\pkg{bridgestan}:}~\log\ev_0 &= -156.4776 \pm 0.0005~\text{[1.1 million evaluations]}
\end{align}
and for the evidence for the alternative,
\begin{align}
\text{\polystan{}:}~\log\ev_1 &= -156.6 \pm 0.2~\text{[5.2 million evaluations]}\\
\text{\pkg{bridgestan}:}~\log\ev_1 &= -156.718 \pm 0.005~\text{[3.4 million evaluations]}
\end{align}
These evidences resulted in Bayes factors that favored the null,
\begin{align}
\text{\polystan{}:}~B_{01} &= 1.3\pm 0.3\\
\text{\pkg{bridgestan}:}~B_{01} &= 1.271 \pm 0.006
\end{align}
In this example, \polystan{} reported a $p$-value from the insertion index test in \cref{sec:test} of zero. This was due to a plateau in the likelihood function at $\like = 0$. The priors in these models resulted in a non-zero measure for a Bernoulli distribution parameter that was numerically indistinguishable from 1 or from 0.\footnote{The existence of this region may indicate that the priors were poorly chosen.} These resulted in zero likelihood when an observation occurred that was predicted with $p = 0$ or when an observation did not occur that was predicted with $p = 1$. These plateaus may spoil evidence estimates~\citep{Fowlie:2020gfd}, though evidences can be resummed taking them into account using e.g., \pkg{anesthetic}~\citep{anesthetic}. 

\twosubexample{Change-point}{disaster}

This a change-point model for the number of coal mining disasters per year in the UK between 1851 and 1962 using data from \cite{jarrett1979note}, following the example in \citet[\href{https://mc-stan.org/docs/stan-users-guide/latent-discrete.html\#model-with-latent-discrete-parameter}{Model with latent discrete parameter}]{stan}. The number of disasters per year is modeled by a Poisson distribution. The rate parameter, though, changes abruptly at an unknown year between 1851 and 1962. The rate parameters before and after the change are assigned exponential priors and the change-point follows a discrete uniform distribution.

\begin{figure}
    \centering
    \includegraphics[keepaspectratio=true,height=0.4\linewidth]{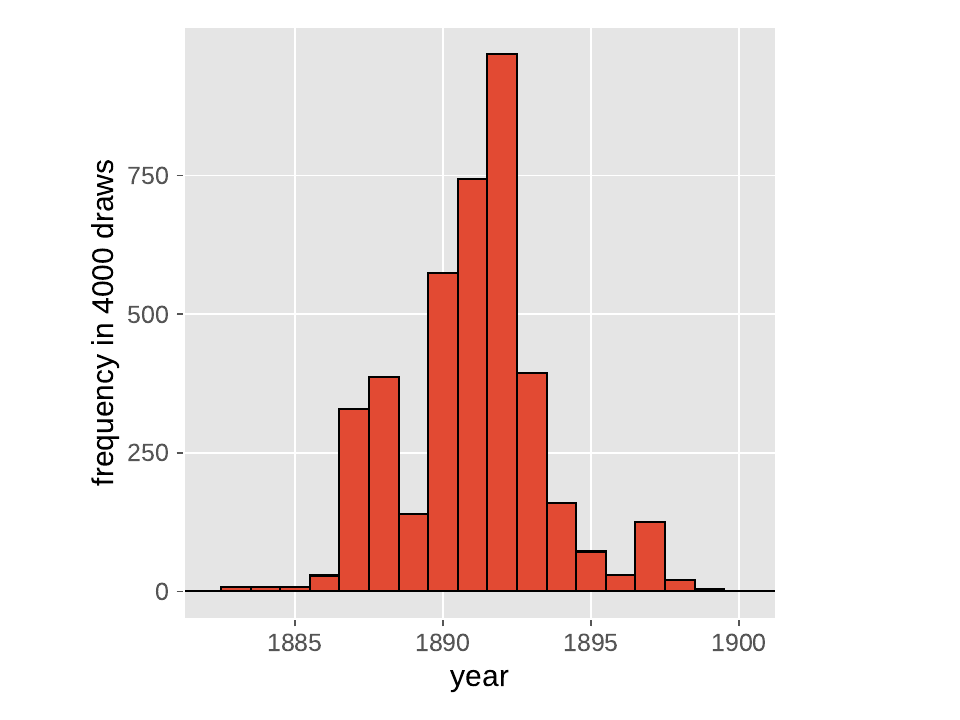}
    \includegraphics[keepaspectratio=true,height=0.4\linewidth]{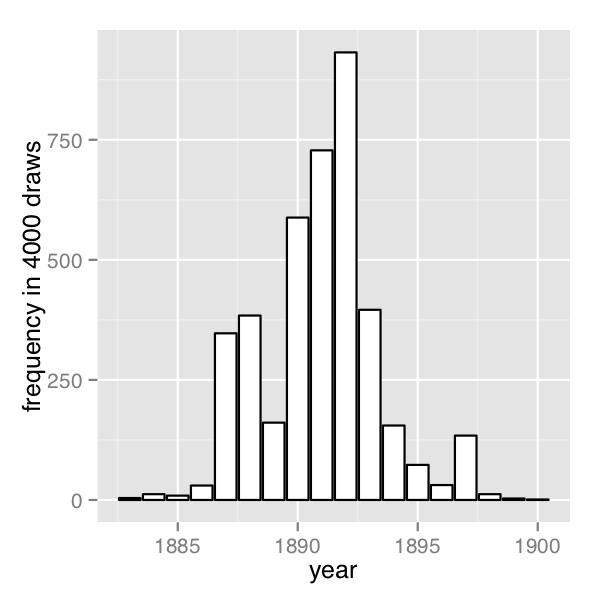}
    \caption{Posterior distribution of a discrete change-point parameter in the coal mining disasters data~\citep{jarrett1979note} found by \polystan{} (left, red) and from \citet[\href{https://mc-stan.org/docs/stan-users-guide/latent-discrete.html\#posterior-distribution-of-the-discrete-change-point}{Posterior distribution of the change point}]{stan} (right, white).}
    \label{fig:disaster}
\end{figure}

This is simple to program using functionality from \code[Stan]{polystan.stanfunctions}:
\ifarxiv%
\begin{lstlisting}[language=stan, numbers=left, title=\texttt{polystan/examples/disaster.stan}]
functions {
  #include polystan.stanfunctions
}
data {
  real<lower=0> r_e;
  real<lower=0> r_l;
  int<lower=1> T;
  array[T] int<lower=0> D;
}
parameters {
  vector<lower=0, upper=1>[3] x;
}
transformed parameters {
  real e = exponential_prior(x[1], r_e);
  real l = exponential_prior(x[2], r_l);
}
model {
  int s = flat_prior(x[3], T);
  vector[T] rate = rep_vector(l, T);
  rate[1 : s] = rep_vector(e, s);
  D ~ poisson(rate);
}
generated quantities {
  int save_s = flat_prior(x[3], T);
}
\end{lstlisting}
\else%
\begin{CodeChunk}
\verbatiminput{disaster.stan} 
\end{CodeChunk}
\fi%
We do not need to program the marginalization of the change-point by hand.\footnote{Unfortunately, we must repeat the computation of the change-point in the \code[Stan]{generated quantities} block so that it is saved to disk, as integer parameters cannot be defined in the \code[Stan]{transformed parameters} block and are not saved in the \code[Stan]{model} block.} We find the change point distribution in \cref{fig:disaster} closely matches that in \citet[\href{https://mc-stan.org/docs/stan-users-guide/latent-discrete.html\#posterior-distribution-of-the-discrete-change-point}{Posterior distribution of the change point}]{stan}.

However, in general, care should be taken around plateaus in the likelihood function for discrete parameters. If there are plateaus, one should consider specialized software e.g., \pkg{anesthetic}~\citep{anesthetic} to correctly analyze NS results; see \cite{Fowlie:2020gfd} for further discussion.

\section{Summary} \label{sec:summary}

We introduced \polystan{} --- a nested sampling inference engine for \proglang{Stan}. \polystan{} couples \proglang{Stan} to the \pkg{PolyChord} nested sampling algorithm using \pkg{bridgestan}. By replacing points one by one, nested sampling evolves a set of live points upwards in likelihood. \pkg{PolyChord} uses slice sampling to draw replacements from the likelihood constrained prior. As the interface mirrors \pkg{cmdstan}, it should be familiar to  \proglang{Stan} users and developers, and provides a black-box method for computing marginal likelihoods. 

We demonstrated \polystan{} on a variety of example problems including generalized linear mixed models, comparing results against the \proglang{Stan} and \pkg{bridgesampling} implementations of HMC and bridge sampling. The examples showed that \pkg{PolyChord} handled multi-modal problems, whereas HMC and thus bridge sampling failed, and models with discrete parameters. On the other hand, when they worked, the combination of HMC and bridge sampling generally resulted in more precise estimates for a similar computational cost. These results, however, depend on choices of hyperparameters for the computer codes.

In summary, \polystan{} introduces a new user-base to nested sampling algorithms, opens up degenerate and multi-modal problems that were previously challenging in \proglang{Stan}, and provides a robust method to compute the marginal likelihood.

\section*{Acknowledgments}

AF was supported by RDF-22-02-079. 

\bibliography{refs}

\begin{thebibliography}{28}
\newcommand{\enquote}[1]{``#1''}
\providecommand{\natexlab}[1]{#1}
\providecommand{\url}[1]{\texttt{#1}}
\providecommand{\urlprefix}{URL }
\expandafter\ifx\csname urlstyle\endcsname\relax
  \providecommand{\doi}[1]{doi:\discretionary{}{}{}#1}\else
  \providecommand{\doi}{doi:\discretionary{}{}{}\begingroup \urlstyle{rm}\Url}\fi
\providecommand{\eprint}[2][]{\url{#2}}

\bibitem[{Aitken and Akman(2013)}]{Aitken2013}
Aitken S, Akman OE (2013).
\newblock \enquote{Nested sampling for parameter inference in systems biology: application to an exemplar circadian model.}
\newblock \emph{BMC Syst. Biol.}, \textbf{7} (1), 72.
\newblock \doi{10.1186/1752-0509-7-72}.

\bibitem[{Ashton \emph{et~al.}(2022)}]{Ashton:2022grj}
Ashton G, \emph{et~al.} (2022).
\newblock \enquote{{Nested sampling for physical scientists}.}
\newblock \emph{Nature}, \textbf{2}.
\newblock \doi{10.1038/s43586-022-00121-x}.
\newblock \eprint{2205.15570}.

\bibitem[{Betancourt(2017)}]{Betancourt:2017ebh}
Betancourt M (2017).
\newblock \enquote{{A Conceptual Introduction to Hamiltonian Monte Carlo}.}
\newblock \eprint{1701.02434}.

\bibitem[{Buchner(2021)}]{Buchner:2021kpm}
Buchner J (2021).
\newblock \enquote{{Nested sampling methods}.}
\newblock \emph{Stat. Surv.}, \textbf{17}.
\newblock \doi{10.1214/23-ss144}.
\newblock \eprint{2101.09675}.

\bibitem[{Dickey(1971)}]{10.1214/aoms/1177693507}
Dickey JM (1971).
\newblock \enquote{{The Weighted Likelihood Ratio, Linear Hypotheses on Normal Location Parameters}.}
\newblock \emph{Ann. Math. Stat.}, \textbf{42} (1), 204 -- 223.
\newblock \doi{10.1214/aoms/1177693507}.

\bibitem[{Fowlie \emph{et~al.}(2020)Fowlie, Handley, and Su}]{Fowlie:2020mzs}
Fowlie A, Handley WJ, Su L (2020).
\newblock \enquote{{Nested sampling cross-checks using order statistics}.}
\newblock \emph{Mon. Not. Roy. Astron. Soc.}, \textbf{497} (4), 5256--5263.
\newblock \doi{10.1093/mnras/staa2345}.
\newblock \eprint{2006.03371}.

\bibitem[{Fowlie \emph{et~al.}(2021)Fowlie, Handley, and Su}]{Fowlie:2020gfd}
Fowlie A, Handley WJ, Su L (2021).
\newblock \enquote{{Nested sampling with plateaus}.}
\newblock \emph{Mon. Not. Roy. Astron. Soc.}, \textbf{503} (1), 1199--1205.
\newblock \doi{10.1093/mnras/stab590}.
\newblock \eprint{2010.13884}.

\bibitem[{Gronau \emph{et~al.}(2020)Gronau, Singmann, and Wagenmakers}]{Gronau2020}
Gronau QF, Singmann H, Wagenmakers EJ (2020).
\newblock \enquote{\pkg{bridgesampling}: An \proglang{R} Package for Estimating Normalizing Constants.}
\newblock \emph{J. Stat. Softw.}, \textbf{92} (10).
\newblock \doi{10.18637/jss.v092.i10}.

\bibitem[{Handley(2019)}]{anesthetic}
Handley WJ (2019).
\newblock \enquote{\pkg{anesthetic}: nested sampling visualisation.}
\newblock \emph{J. Open Source Softw.}, \textbf{4} (37), 1414.
\newblock \doi{10.21105/joss.01414}.

\bibitem[{Handley \emph{et~al.}(2015{\natexlab{a}})Handley, Hobson, and Lasenby}]{Handley:2015fda}
Handley WJ, Hobson MP, Lasenby AN (2015{\natexlab{a}}).
\newblock \enquote{{\pkg{PolyChord}: nested sampling for cosmology}.}
\newblock \emph{Mon. Not. Roy. Astron. Soc.}, \textbf{450} (1), L61--L65.
\newblock \doi{10.1093/mnrasl/slv047}.
\newblock \eprint{1502.01856}.

\bibitem[{Handley \emph{et~al.}(2015{\natexlab{b}})Handley, Hobson, and Lasenby}]{Handley:2015vkr}
Handley WJ, Hobson MP, Lasenby AN (2015{\natexlab{b}}).
\newblock \enquote{{\pkg{polychord}: next-generation nested sampling}.}
\newblock \emph{Mon. Not. Roy. Astron. Soc.}, \textbf{453} (4), 4385--4399.
\newblock \doi{10.1093/mnras/stv1911}.
\newblock \eprint{1506.00171}.

\bibitem[{Hunter(2007)}]{Hunter:2007}
Hunter JD (2007).
\newblock \enquote{\pkg{Matplotlib}: A 2D graphics environment.}
\newblock \emph{Computing in Science \& Engineering}, \textbf{9} (3), 90--95.
\newblock \doi{10.1109/MCSE.2007.55}.

\bibitem[{Janzen \emph{et~al.}(2000)Janzen, Tucker, and Paukstis}]{Janzen2000}
Janzen FJ, Tucker JK, Paukstis GL (2000).
\newblock \enquote{Experimental analysis of an early life-history stage: Selection on size of hatchling turtles.}
\newblock \emph{Ecology}, \textbf{81} (8), 2290–2304.
\newblock ISSN 0012-9658.
\newblock \doi{10.1890/0012-9658(2000)081[2290:eaoael]2.0.co;2}.

\bibitem[{Jarrett(1979)}]{jarrett1979note}
Jarrett RG (1979).
\newblock \enquote{A note on the intervals between coal-mining disasters.}
\newblock \emph{Biometrika}, \textbf{66} (1), 191--193.

\bibitem[{Jeffreys(1939)}]{Jeffreys:1939xee}
Jeffreys H (1939).
\newblock \emph{{The Theory of Probability}}.
\newblock Oxford University Press.
\newblock ISBN 978-0-19-850368-2.

\bibitem[{Kass and Raftery(1995)}]{Kass01061995}
Kass RE, Raftery AE (1995).
\newblock \enquote{Bayes Factors.}
\newblock \emph{J. Am. Stat. Assoc.}, \textbf{90} (430), 773--795.
\newblock \doi{10.1080/01621459.1995.10476572}.

\bibitem[{Kumar \emph{et~al.}(2019)Kumar, Carroll, Hartikainen, and Martin}]{arviz_2019}
Kumar R, Carroll C, Hartikainen A, Martin O (2019).
\newblock \enquote{\pkg{ArviZ} a unified library for exploratory analysis of Bayesian models in \proglang{Python}.}
\newblock \emph{J. Open Source Softw.}, \textbf{4} (33), 1143.
\newblock \doi{10.21105/joss.01143}.

\bibitem[{Massey(1951)}]{Massey1951}
Massey FJ (1951).
\newblock \enquote{The Kolmogorov-Smirnov Test for Goodness of Fit.}
\newblock \emph{J. Am. Stat. Assoc.}, \textbf{46} (253), 68–78.
\newblock \doi{10.1080/01621459.1951.10500769}.

\bibitem[{Meng and Wong(1996)}]{meng1996simulating}
Meng XL, Wong WH (1996).
\newblock \enquote{Simulating ratios of normalizing constants via a simple identity: a theoretical exploration.}
\newblock \emph{Stat. Sin.}, pp. 831--860.
\newblock \urlprefix\url{https://www.jstor.org/stable/24306045}.

\bibitem[{Neal(2003)}]{Neal2003}
Neal RM (2003).
\newblock \enquote{Slice sampling.}
\newblock \emph{Ann. Stat.}, \textbf{31} (3).
\newblock \doi{10.1214/aos/1056562461}.
\newblock \eprint{physics/0009028}.

\bibitem[{Neal(2011)}]{Neal:2011mrf}
Neal RM (2011).
\newblock \emph{{Handbook of Markov Chain Monte Carlo}}.
\newblock Chapman and Hall/CRC.
\newblock \doi{10.1201/b10905}.
\newblock \eprint{1206.1901}.

\bibitem[{Park(2021)}]{park2021sampling}
Park J (2021).
\newblock \enquote{Sampling from high-dimensional, multimodal distributions using automatically tuned, tempered Hamiltonian Monte Carlo.}
\newblock \eprint{2111.06871}.

\bibitem[{Roualdes \emph{et~al.}(2023)Roualdes, Ward, Carpenter, Seyboldt, and Axen}]{Roualdes2023}
Roualdes EA, Ward B, Carpenter B, Seyboldt A, Axen SD (2023).
\newblock \enquote{\pkg{BridgeStan}: Efficient in-memory access to the methods of a \proglang{Stan} model.}
\newblock \emph{J. Open Source Softw.}, \textbf{8} (87), 5236.
\newblock \doi{10.21105/joss.05236}.

\bibitem[{Schreiner \emph{et~al.}()}]{cli11}
Schreiner H, \emph{et~al.}
\newblock \enquote{{\pkg{CLI11}: Command line parser for \proglang{C++11}}.}
\newblock \urlprefix\url{https://github.com/CLIUtils/CLI11}.

\bibitem[{Skilling(2004)}]{Skilling2004}
Skilling J (2004).
\newblock \enquote{Nested Sampling.}
\newblock In \emph{AIP Conference Proceedings}, volume 735, p. 395–405. AIP.
\newblock \doi{10.1063/1.1835238}.

\bibitem[{Skilling(2006)}]{10.1214/06-BA127}
Skilling J (2006).
\newblock \enquote{{Nested sampling for general Bayesian computation}.}
\newblock \emph{Bayesian Anal.}, \textbf{1} (4), 833 -- 859.
\newblock \doi{10.1214/06-BA127}.

\bibitem[{{\proglang{Stan} Development Team}(2024)}]{stan}
{\proglang{Stan} Development Team} (2024).
\newblock \enquote{{\proglang{Stan} Modeling Language User's Guide and Reference Manual}.}
\newblock \urlprefix\url{https://mc-stan.org}.

\bibitem[{Vehtari \emph{et~al.}(2021)Vehtari, Gelman, Simpson, Carpenter, and Bürkner}]{Vehtari_2021}
Vehtari A, Gelman A, Simpson D, Carpenter B, Bürkner PC (2021).
\newblock \enquote{{Rank-Normalization, Folding, and Localization: An Improved $\hat R$ for Assessing Convergence of MCMC}.}
\newblock \emph{Bayesian Anal.}, \textbf{16} (2).
\newblock \doi{10.1214/20-ba1221}.
\newblock \eprint{1903.08008}.

\end{thebibliography}

\clearpage
\appendix

\section{Provided transform functions}\label{app:prior_transforms}

Transforms for common choices of prior are accessible by including the \code[stan]{polystan.stanfunctions} header:
\begin{CodeChunk}
\begin{CodeInput}[Stan]
functions {
  #include polystan.stanfunctions
}
\end{CodeInput}
\end{CodeChunk}
\polystan{} includes the path to this header in the build. The naming scheme matches \proglang{Stan} as closely as possible, e.g., \code[stan]{std_normal_prior}. The signatures are the draw from a uniform distribution to be transformed, followed by any parameters that specify the distribution.

These transformatons can be checked using the \code{polystan/contrib/inspect_priors.py} script, which builds and runs the \code{polystan/examples/priors.stan} code and plots the results. 

\ifarxiv%
\begin{colormath}
\begin{lstlisting}[language=stan, mathescape, numbers=left, title=\texttt{polystan/stanfunctions/polystan.stanfunctions}]
// inverse transform sampling from $x \sim \mathcal{U}(0, 1)$

real flat_prior(real x, real a, real b) {
  // if $x \sim \mathcal{U}(0, 1)$ then $y \sim \mathcal{U}(a, b)$
  return (b - a) * x + a;
}

vector flat_prior(vector x, real a, real b) {
  // if $x \sim \mathcal{U}(0, 1)$ then $y \sim \mathcal{U}(a, b)$
  return (b - a) * x + a;
}

real log_prior(real x, real a, real b) {
  // if $x \sim \mathcal{U}(0, 1)$ then $\log y \sim \mathcal{U}(\log a, \log b)$
  return a * exp(log(b / a) * x);
}

real std_normal_prior(real x) {
  // if $x \sim \mathcal{U}(0, 1)$ then $y \sim \mathcal{N}(0, 1)$
  return inv_Phi(x);
}

vector std_normal_prior(vector x) {
  // if $\vec x \sim \mathcal{U}(0, 1)$ then $\vec y \sim \mathcal{N}(0, 1)$
  return inv_Phi(x);
}

real half_std_normal_prior(real x) {
  // if $x \sim \mathcal{U}(0, 1)$ then $y \sim \mathcal{N}_{>0}(0, 1)$
  return std_normal_prior(flat_prior(x, 0.5, 1.));
}

real normal_prior(real x, real mu, real sigma) {
  // if $x \sim \mathcal{U}(0, 1)$ then $y \sim \mathcal{N}(\mu, \sigma^2)$
  return inv_Phi(x) * sigma + mu;
}

vector normal_prior(vector x, real mu, real sigma) {
  // if $\vec x \sim \mathcal{U}(0, 1)$ then $\vec y \sim \mathcal{N}(\mu, \sigma^2)$
  return inv_Phi(x) * sigma + mu;
}

real half_normal_prior(real x, real mu, real sigma) {
  // if $x \sim \mathcal{U}(0, 1)$ then $y \sim \mathcal{N}_{>\mu}(\mu, \sigma^2)$
  return normal_prior(flat_prior(x, 0.5, 1.), mu, sigma);
}

vector multi_normal_prior(vector x, vector mu, matrix Sigma) {
  // if $\vec x \sim \mathcal{U}(0, 1)$ then $\vec y \sim \mathcal{N}(\vec \mu, \vec \Sigma)$
  return multi_normal_cholesky_prior(x, mu, cholesky_decompose(Sigma));
}

vector multi_normal_cholesky_prior(vector x, vector mu, matrix L) {
  // if $\vec x \sim \mathcal{U}(0, 1)$ then $\vec y \sim \mathcal{N}(\vec \mu, \vec \Sigma = \vec L \vec L^T)$
  return L * inv_Phi(x) + mu;
}

real log10normal_prior(real x, real mu, real sigma) {
  // if $x \sim \mathcal{U}(0, 1)$ then $\log_{10} y \sim \mathcal{N}(\mu, \sigma^2)$
  return exp(log(10.) * normal_prior(x, mu, sigma));
}

real lognormal_prior(real x, real mu, real sigma) {
  // if $x \sim \mathcal{U}(0, 1)$ then $\log y \sim \mathcal{N}(\mu, \sigma^2)$
  return exp(normal_prior(x, mu, sigma));
}

real beta_prior(real x, real alpha, real beta) {
  // if $x \sim \mathcal{U}(0, 1)$ then $y \sim \text{Beta}(\alpha, \beta)$
  return inv_inc_beta(alpha, beta, x);
}

real beta_prime_prior(real x, real alpha, real beta) {
  // if $x \sim \mathcal{U}(0, 1)$ then $y \sim \text{Beta}'(\alpha, \beta)$
  real p = inv_inc_beta(alpha, beta, x);
  return p / (1. - p);
}

real dagum_prior(real x, real p, real a, real b) {
  // if $x \sim \mathcal{U}(0, 1)$ then $y \sim \text{Dag}(p, a, b)$
  return b * pow(pow(x, -1. / p) - 1., -1. / a);
}

real exponential_prior(real x, real lambda) {
  // if $x \sim \mathcal{U}(0, 1)$ then $y \sim \text{Exp}(\lambda)$
  return -log(1. - x) / lambda;
}

real cauchy_prior(real x, real x0, real gamma) {
  // if $x \sim \mathcal{U}(0, 1)$ then $y \sim \text{Cauchy}(x_0, \gamma)$
  return x0 + gamma * tan(pi() * (x - 0.5));
}

real half_cauchy_prior(real x, real x0, real gamma) {
  // if $x \sim \mathcal{U}(0, 1)$ then $y \sim \text{Cauchy}_{>x_0}(x_0, \gamma)$
  return cauchy_prior(flat_prior(x, 0.5, 1.), x0, gamma);
}

int categorical_prior(real x, vector p) {
  // if $x \sim \mathcal{U}(0, 1)$ then $y \sim \text{Categorical}(p)$
  int N = size(p);
  vector[N] cmf = cumulative_sum(p);
  for (i in 1 : N - 1) {
    if (x < cmf[i]) {
      return i;
    }
  }
  return N;
}

int flat_prior(real x, int N) {
  // if $x \sim \mathcal{U}(0, 1)$ then $y \sim \mathcal{U}(1, N)$
  for (i in 1 : N - 1) {
    if (x < i * 1. / N) {
      return i;
    }
  }
  return N;
}
\end{lstlisting}
\end{colormath}
\else%
\begin{CodeChunk}
\verbatiminput{polystan.stanfunctions} 
\end{CodeChunk}
\fi%

\end{document}